# VLBI measurement of the vector baseline between geodetic antennas at Kokee Park Geophysical Observatory, Hawaii

A.E. Niell[1], J.P. Barrett[1], R.J. Cappallo[1], B.E. Corey[1], P. Elosegui[1,2], D. Mondal[1], G. Rajagopalan[1], C.A. Ruszczyk[1], M.A. Titus[1]

[1]Massachusetts Institute of Technology Haystack Observatory, Westford, MA, USA

[2]Institute of Marine Sciences, ICM-CSIC, Barcelona, Spain

Corresponding author: A.E. Niell (aniell@mit.edu)

ORCID information

| | |
|---|---|
| Barrett: | 0000-0002-9290-0764 |
| Cappallo: | 0000-0002-8460-1844 |
| Corey: | 0000-0001-7809-7357 |
| Elosegui: | 0000-0002-4120-7855 |
| Mondal | 0000-0003-0708-6575 |
| Niell | 0000-0001-5267-4005 |
| Rajagopalan: | 0000-0003-0224-3017 |
| Ruszczyk: | 0000-0001-7278-9707 |
| Titus: | 0000-0002-3423-4505 |

## Abstract

We measured the components of the 31-m-long vector between the two Very-Long-Baseline Interferometry (VLBI) antennas at the Kokee Park Geophysical Observatory (KPGO), Hawaii, with approximately 1 mm precision using phase-delay observables from dedicated VLBI observations in 2016 and 2018. The two KPGO antennas are the 20 m legacy VLBI antenna and the 12 m VLBI Global Observing System (VGOS) antenna. Independent estimates of the vector between the two antennas were obtained by the National Geodetic Survey (NGS) using standard optical surveys in 2015 and 2018. The uncertainties of the latter survey were 0.3 and 0.7 mm in the horizontal and vertical components of the baseline, respectively. We applied corrections to the measured positions for the varying thermal deformation of the antennas on the different days of the VLBI and survey measurements, which can amount to 1 mm, bringing all results to a common reference temperature. The difference between the VLBI and survey results are 0.2 ± 0.4 mm, -1.3 ± 0.4 mm, and 0.8 ± 0.8 mm in the East, North, and Up topocentric components, respectively. We also estimate that the Up component of the baseline may suffer from systematic errors due to gravitational deformation and uncalibrated instrumental delay variations at the 20 m antenna that may reach ±10 mm and -2 mm, respectively, resulting in an accuracy uncertainty on the order of 10 mm for the relative heights of the antennas. Furthermore, possible tilting of the 12 m antenna increases the uncertainties in the differences in the horizontal components to 1.0 mm. These results bring into focus the importance of (1) correcting to a common reference temperature the measurements of the reference points of all geodetic instruments within a site, (2) obtaining measurements of the gravitational deformation of all antennas, and (3) monitoring local motions of the geodetic instruments. These results have significant implications for the accuracy of global reference frames that require accurate local ties between geodetic instruments, such as the International Terrestrial Reference Frame (ITRF).

Keywords       Geodetic VLBI; reference frames; ITRF; Global Geodetic Observing System; core sites; local vector ties; phase-delay VLBI; antenna thermal deformation

# 1 Introduction

A quantitative description of the shape of the Earth and how it changes with time, as provided by the International Terrestrial Reference Frame (ITRF, see, e.g., Altamimi et al. 2016), is fundamental to both the understanding of the Earth's structure and the functioning of society. Construction of the ITRF incorporates extraordinarily precise measurements by instruments both on the surface of the Earth and in space. The four techniques contributing to the ITRF are currently the Global Navigation Satellite System (GNSS), Satellite Laser Ranging (SLR), Very Long Baseline Interferometry (VLBI), and Doppler Orbitography and Radiopositioning Integrated by Satellite (DORIS). GNSS and DORIS rely on measuring the distances to Earth-orbiting satellites using radio waves. SLR operates on the same principles but the signals are pulses of light. VLBI is a technique that measures the difference in arrival times, at two antennas, of the radio signals from distant objects in the Universe. Each of these techniques determines the relative positions of the same type of instruments on the ground, e.g., GNSS-to-GNSS, with an accuracy of typically a few millimeters (e.g., Herring et al. 2016). However, the results of the different systems must be combined to obtain the highest accuracy and to relate the points on the surface of the Earth whose positions are not determined by the same technique.

An essential component in the construction of a global reference frame is the set of vector connections (ties) among the geodetic instruments of the four techniques at a common site, including ties among instruments of the same type (e.g., multiple GNSS or VLBI antennas) (Ray and Altamimi 2005, Altamimi et al. 2016, Glaser et al. 2019).

The next-generation geodetic VLBI system, designated VLBI Global Observing System (VGOS), will deliver greater geodetic precision than that provided by the legacy VLBI antennas (e.g., Niell et al. 2018, and references therein). However, the coordinates of a network of VGOS antennas operating independently to obtain the best precision will be independent of the legacy network frame until the networks are tied together. Therefore, the new VGOS antennas must be integrated with both the legacy VLBI network and the other techniques.

The three most precise ways to relate the VGOS antennas to the other systems (considering legacy VLBI as a separate system) are: 1) inter-technique: do an optical survey that includes the VGOS antenna(s) and the co-located VLBI, GPS, SLR, or DORIS systems (e.g., Carter et al. 1979; Erickson and Breidenbach 2019); a limitation to the accuracy of this method is the uncertainty in the relation of the optical reference points to the electromagnetic reference point; 2) intra-technique by network: include the VGOS antennas in the legacy network observations (called mixed-mode observing) to incorporate those VGOS antennas that are not near (greater than a few tens of kilometers) to a legacy antenna; this is the only means for directly tying a VGOS antenna that is remote from any legacy antenna into the legacy frame; 3) intra-technique by antenna: use VLBI to directly tie those VGOS and legacy antennas that are co-located at a site (e.g., Herring 1992) either by including both antennas observing together in global legacy sessions or by scheduling separate tie sessions for all of the pairs or triples of antennas at each co-location site. For either case the significant reduction in delay uncertainty that is readily achievable on very short baselines by using phase-delay instead of group-delay observations enables relative positional precision on the millimeter to sub-millimeter level.

A fourth method is to utilize GNSS systems attached to the VLBI antenna to estimate, from a series of measurements made while the VLBI antenna moves around the sky, the location of the apparent center of motion (invariant point) of the VLBI antenna (Ning et al. 2015, and references therein) relative to a nearby reference GNSS antenna. This method is limited by the inherent indeterminacy of the offsets between the electrical and physical reference points of both the GNSS and VLBI antennas, including for GNSS the sensitivity to the minimum observed elevation. However, the method does offer a convenient and efficient means to monitor the tie vector on a regular basis during operational VLBI observations. Furthermore, it could be extended to multiple co-located VLBI antennas.

The first of the three primary options is the only practical way (at this time) to tie all of the disparate techniques to a common point at a co-located site (Altamimi et al. 2016). It also provides an independent measurement for comparison with the intra-technique-by-antenna tie (third option). The second and third options are for tying the VGOS and legacy VLBI systems in a common frame.

For the latter two options, a VGOS antenna must operate as though it were a legacy antenna in terms of frequency coverage and schedule (the sequence of observations within a session), and thus the limitations of the legacy systems (lower antenna slew rates and limited frequency coverage) adversely impact the benefits for which the VGOS systems were designed. As a consequence, trying to determine the position of a co-located VGOS antenna in the legacy frame through participation in the legacy sessions, but without the co-located legacy antenna (second option), is much less efficient than the direct tie of the third option. Without the pairing of the co-located legacy and VGOS antennas in a session, the position uncertainty of the VGOS antenna will not be better than a legacy antenna of comparable sensitivity, which is on the order of 1–4 mm in local horizontal and 4–7 mm in local vertical for a typical 24-hr session. To obtain sub-millimeter precision in the reference frame this way, it will take global sessions spread over a year or more at the current rate of observing of once or twice per week.

The most important benefit of the second and third options is that the measurements directly relate the radio reference points of the co-located antennas. For VLBI, the reference point is the intersection of axes (indicated in Fig. 4) and may not be a physically accessible point on the antenna structure, while the optical and GNSS ties require measurements to physical points on the antenna that may not coincide with the electrical properties. In addition, the radio-tie observations can be made much more quickly and easily than an optical tie, thus allowing more frequent repetition to evaluate possible changes with time. The only shortcoming is that such a tie is useful only for the VLBI-to-VLBI tie, not to any of the other space geodetic instruments or to the site reference marker.

Between the two possibilities for carrying out the third option, the dedicated local tie sessions offer several potential advantages over participation in global legacy sessions. Not being constrained by the requirement of having available one or more distant antennas for each scan, the co-located antennas can attain greater sky coverage and a higher temporal density. This is achieved by using shorter scans for the same signal-to-noise ratio (SNR) requirement by virtue of sources being generally stronger on shorter baselines, as well as by allowing a lower minimum SNR because of the higher precision of the phase-delay observable.

The legacy network has been in operation for more than three decades. By virtue of this long record, the legacy stations have estimates of both position and velocity with high precision (e.g., Altamimi et al. 2016). Tying a VGOS antenna to a co-located legacy antenna is the best of both worlds in that high-precision velocity estimates of the legacy antennas are effectively incorporated into the higher precision position estimates of VGOS.

In this paper we present the results of VLBI measurements to obtain a vector tie between the adjacent legacy and VGOS geodetic VLBI antennas at Kokee Park Geophysical Observatory (KPGO), Kauai, Hawaii, using phase-delay observations, and compare that tie to direct optical surveys.

High-precision phase-delay measurements of the vector baseline between co-located VLBI antennas have been reported previously for the Haystack 37 m and Westford 18 m antennas in Westford, MA, USA (Rogers et al. 1978; Carter et al. 1980; Herring 1992). The main point of the Carter et al. paper was to demonstrate the agreement with an optical survey of the baseline, including the importance of correcting for gravitational deformation of the 37 m antenna. Herring (1992) analyzed several years of data to validate the constancy of the vector separation of the two antennas as required for a stable global reference frame. The measurements reported here differ from those earlier results in providing the first documentation of the VLBI tie between the VGOS and legacy antennas at KPGO, as well as describing the mixed-mode observations with those antennas and analysis of the results.

The paper is organized as follows. In Section 2 we describe the two antennas and their hardware configurations, clarifying the differences between the legacy and VGOS systems. Sections 3 and 4 contain a description of the observation planning and procedures, the special requirements of the mixed-mode (i.e., legacy and VGOS) data acquisition, and the correlation and post-correlation processing. The geodetic analysis of the derived VLBI delay observables is covered in section 5. Effects that change the estimated vector baseline, such as thermal deformation of the antennas, are discussed in Section 6, and the corrected results are presented in Section 7. In Section 8 these results are compared to optical surveys of the same vector baseline. Section 9 contains a summary of the main points and discusses improvements that can be made in future measurements of this type.

## 2 Experimental setup

### 2.1 Differences between legacy and VGOS antennas and instrumentation

The antennas at KPGO are examples of the two generations of geodetic VLBI networks (legacy and VGOS) whose coordinates must be accurately related to achieve the best terrestrial reference frame (Fig. 1).

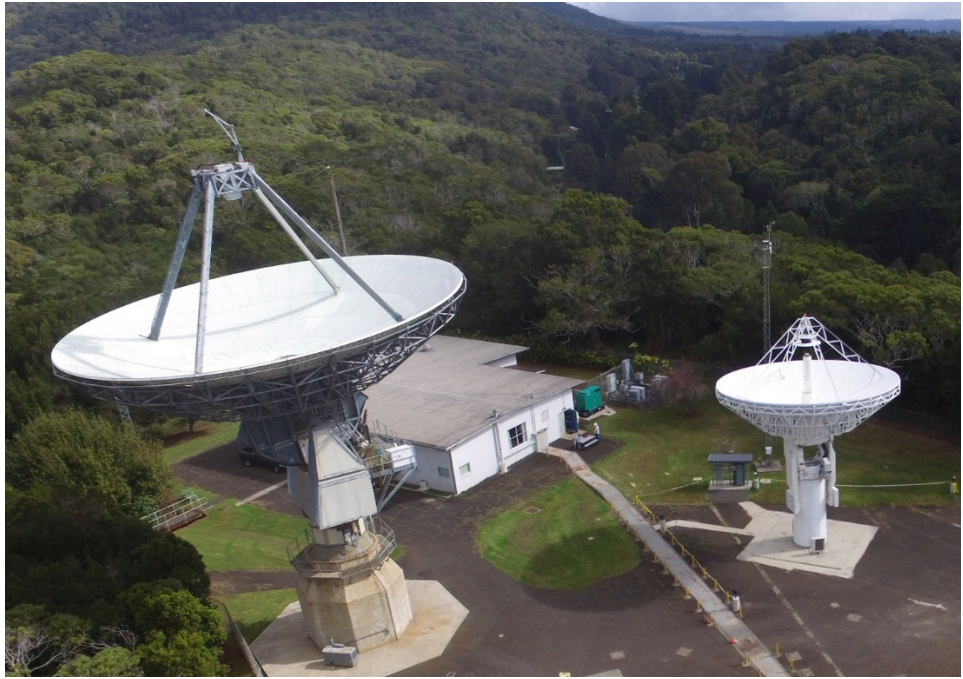

**Fig. 1** View of the Kokee Park Geophysical Observatory (KPGO) showing the 20 m legacy antenna (left) and 12 m VGOS antenna (right); source Chris Coughlin and Kiah Imai, KPGO

They differ in the size and performance of the antenna structures and in the capability of the instrumentation. The legacy antennas are typically larger than the VGOS antennas, thus in general providing greater sensitivity in the two radio frequency bands they utilize, but they are slower in moving from one radio source to another. The VGOS broadband instrumentation (Niell et al. 2018), on the other hand, covers a much wider frequency range (which includes the legacy bands) and has a much greater data acquisition rate. Taken together these properties give the VGOS systems significantly better delay precision in a given integration time. Even though the instrumentation is completely different for the two systems, it is possible to obtain measurements between them because the frequencies of the legacy system are a subset of the VGOS system. The primary differences between the legacy and VGOS systems are discussed in Niell et al. (2018). We review here only those characteristics that are relevant to the mixed legacy/VGOS observing mode of this study.

## 2.2 Frequencies

The legacy geodetic VLBI systems, also referred to as S/X, including the KPGO 20 m antenna, receive the radio signals at S-band (2.2-2.4 GHz) and X-band (8.2-8.9 GHz) (Rogers et al. 1983). The VGOS systems currently utilize four 512 MHz-wide radio-frequency bands, each of which can be set anywhere within the range approximately 2.2 to 14 GHz (Petrachenko et al. 2009). (A diagram of the main components of the signal chain can be found in Niell et al. (2018).) Therefore, for the KPGO tie sessions, the frequencies selected for reception by the broadband VGOS antenna were adjusted to encompass the S and X bands of the legacy system.

An additional restriction in the selection of the broadband frequencies was that the 8 MHz-wide channels of the signal chain of the 20 m legacy antenna cannot overlap the edges of the 32 MHz-wide channels of the broadband 12 m system. This limitation is imposed by the capabilities of the DiFX correlator (Deller et al. 2011). In designing the observing sessions, it was also desirable to maintain the channel frequencies of the VLBI observing sessions of the International VLBI Service for Geodesy and Astrometry (IVS), in particular the regularly scheduled weekly IVS 24-hour R1 and R4 series of geodetic VLBI sessions.

With these restrictions, compatibility could be obtained only with the eight upper-sideband X-band channels and five of the six S-band channels. For X-band, two of the four VGOS bands were needed to cover the full legacy X-band since each VGOS band is fully receptive to only 480 MHz, a limitation of the RDBE-G digital back end (Niell et al. 2010). The S-band channels are covered by one VGOS band.

The channels of the fourth VGOS band were kept at the frequencies used for the second band of the operational VGOS observations, which spans 5.27 to 5.75 GHz, in anticipation that this set of frequencies will be used for upcoming S/X-VGOS observations, such as the R1s and R4s. The data from eight channels of each of the four bands were recorded at the VGOS antennas in order to obtain the best delay precision and sensitivity on the VGOS-only baseline (see Section 3). The effective local oscillator (LO) frequencies that were used are given in Table 1.

**Table 1** Frequency channels. The listed frequencies are the effective total local oscillator (LO) frequencies. The VGOS channels are 32 MHz wide, and the S/X channels are 8 MHz. All VGOS channels are lower sideband (LSB) and the S and X channels are upper sideband (USB) unless followed by the letter 'L' (lower sideband). The position of each S/X entry relative to the VGOS entries indicates which VGOS channel the S/X channel falls within. For the lowest and highest X-band frequencies both upper and lower sideband were recorded for the legacy systems and were correlated, but the lower sideband channels were not correlated to the VGOS antennas.

| X-band high VGOS channel | VGOS LO 32 MHz LSB | S/X LO (8 MHz) USB unless noted | | |
|---|---|---|---|---|
| 1 | 9198.4 | - | | |
| 2 | 9166.4 | - | | |
| 6 | 9038.4 | - | | |
| 9 | 8942.4 | 8932.99 | 8932.99L* | 8912.99 |
| 11 | 8878.4 | 8852.99 | | |
| 13 | 8814.4 | - | | |
| 14 | 8782.4 | - | | |
| 15 | 8750.4 | 8732.99 | | |

* not used

| X-band low VGOS channel | VGOS LO 32 MHz LSB | S/X LO (8 MHz) USB unless noted | |
|---|---|---|---|
| 1 | 8686.4 | - | |
| 2 | 8654.4 | - | |
| 4 | 8590.4 | - | |
| 6 | 8526.4 | 8512.99 | |
| 11 | 8366.4 | 8352.99 | |
| 13 | 8302.4 | - | |
| 14 | 8270.4 | 8252.99 | |
| 15 | 8238.4 | 8212.99 | 8212.99L* |

* not used

| C-band VGOS channel | VGOS LO 32 MHz LSB | S/X LO (8 MHz) USB unless noted |
|---|---|---|
| 1 | 5518.4 | - |
| 2 | 5486.4 | - |

| | | | |
|---|---|---|---|
| 4 | 5422.4 | - | |
| 6 | 5358.4 | - | |
| 11 | 5262.4 | - | |
| 13 | 5134.4 | - | |
| 14 | 5102.4 | - | |
| 15 | 5070.4 | - | |

| S-band VGOS channel | VGOS LO 32 MHz LSB | S/X LO (8 MHz) USB unless noted | |
|---|---|---|---|
| 1 | 2702.4 | - | |
| 5 | 2574.4 | - | |
| 8 | 2478.4 | - | |
| 11 | 2382.4 | 2365.99 | 2345.99* |
| 12 | 2350.4 | - | |
| 13 | 2318.4 | 2295.99 | |
| 14 | 2286.4 | 2265.99 | |
| 15 | 2254.4 | 2245.99 | 2225.99 |

\* not used

================================================================================

## 2.3 X-band only for the tie

Geodetic VLBI systems span several GHz in order to be able to estimate and remove the dispersive effect of the charged particles along the lines of sight from the radio source to the antennas. For intercontinental baselines the error in the delay measurement that is incurred if the dispersive effect is not estimated can result in an error in the length of the baseline on the order of several tens of centimeters. However, for baseline lengths up to a few kilometers for which local geodetic ties are needed, the delay error due to neglecting dispersion, which is due primarily to the Earth's ionosphere, is negligible compared to the delay measurement precision (e.g., Rogers et al. 1978). Consequently, observations for the tie sessions can utilize only the X-band data with no loss of accuracy. In fact, the precision is better than if S-band were used to estimate a dispersion because the dispersion correction contributes additional noise to the measurement value and thus increases the delay uncertainty.

## 2.4 Phase calibration

The electrical signal path length can differ between frequency channels due to differences in electrical components and cables in the channels or to frequency-dependent variations in the phase response along a single signal path. The electrical length also varies temporally in a given channel due to changes in temperature or mechanical deformation of those signal chain components. The delay and phase differences among the channels must be removed before the channels are combined in order to restore coherence across the entire frequency range. This is accomplished primarily by tracking the phases of tones injected in the signal chain following the feed (Rogers 1975). These tones are generated from a reference signal by the phase calibrator (hereafter referred to as phasecal) unit. The VGOS system utilizes a 5 MHz reference frequency, and the legacy system derives the tones from a 500 MHz reference frequency, although the phasecal for KOKEE was turned off for these sessions (see section 3). The phases of the tones are extracted in the correlator and used to correct the phases of the cross-correlated signals.

## 2.5 Cable calibration

The benefit of the phase calibration is that the astronomical signal and the phasecal signal are affected the same way by the signal electronics, so removing from the measured VLBI phases the changes observed in the phasecal phases over the common portions of their signal path is a valid correction. On the other hand, any change in the delay of the reference signal from the hydrogen maser to the phasecal generator is not experienced by the astronomical signal, yet it appears as a change in the phasecal delay and thus corrupts the correction to the astronomical delay. Such variation can be due to mechanical deformation or change in the temperature of the cable carrying the reference signal. The delay change due to bending or twisting of the cable is correlated with antenna orientation and, if repeatable, may introduce an artificial apparent change in antenna position. For example, if the cable delay is larger in one azimuth than in the opposite direction, the antenna will appear displaced in that opposite direction.

A unit called the cable calibrator (hereafter referred to as cablecal) has been implemented on legacy antennas from the earliest days of geodetic VLBI to provide measurement of the delay of that reference signal. The measured values are subtracted from the observed delays at the time of the geodetic analysis (see Section 5). Such a system was in place for KOKEE during the tie sessions and the results were utilized. The precision of the cablecal measurements is on the order of a picosecond as evidenced by the peak-to-peak variation of only 1 ps over 30 minutes or more.

For the VGOS systems, a new cable delay measurement system (CDMS) has been developed and was in the process of being implemented on the 12 m antenna at the time of these observations. Unfortunately, it was functional for only part of one session. We therefore used a proxy cablecal approach that we developed which makes use of the phasecal tones to calculate equivalent cable delays (Niell et al. 2018).

## 2.6 Reference frequency and clock initialization

VLBI was developed to allow the astronomical radio signal to be recorded independently at each antenna, thus removing the restriction of having to distribute a common frequency reference to all antennas to provide coherence. The two systems at KPGO are capable of operating independently, each with its own complete signal chain. Although there were two hydrogen masers available, both signal chains used a common 5 MHz reference frequency and 1-pps timing pulse from the same maser. In spite of this commonality, temporal variations in the instrumental delay differences between the two recorded signals were caused by differences in the equipment and in their sensitivities to temperature variations Thus, it was necessary to process the observations as though recorded at entirely separate sites.

## 3 The observations

The first successful VLBI observations with the 12 m antenna at KPGO were made in 2016 February. The progression from that event to operational observing usually would have taken some months as the new observing procedures were learned by station personnel, equipment was validated, and the parameters of the new antenna (e.g., pointing and sensitivity) were better understood. Measurement of the position relative to the 20 m would normally have followed this inaugural period, typically called the commissioning phase. However, replacement of the azimuth bearing of KOKEE had already been scheduled to begin in April, so it was necessary to advance these tie sessions. A possible consequence of the bearing replacement was a change in the position of the intersection of axes. Thus, it was desirable to measure the VLBI baseline at KPGO both before and after the bearing replacement.

Four special VLBI sessions were hurriedly undertaken to measure the position of the new 12 m antenna (hereafter KOKEE12M) relative to the old 20 m antenna (hereafter KOKEE) to have a VLBI tie of the VGOS antenna to the reference point of the legacy S/X antenna that had been in use for almost thirty years.

The intention was to follow with one or more sets of tie sessions after installation of the new bearing in order to measure any change in the position of KOKEE. In these four sessions, the Westford 18 m antenna, also equipped with a broadband feed and signal chain, was added to those observations for which the radio source that was scheduled for the two primary antennas was also visible by Westford and was detectable to either or both of the KPGO antennas (called 'tagalong' scheduling mode).

The sessions, which were designated 'KT' sessions, had planned durations of 1, 6, 24, and 24 hours with the length increasing from an initial test to full operation. However, the third and fourth sessions had some loss of data as indicated in Table 2.

The next opportunity for simultaneous observations with the two KPGO antennas occurred in 2018 December when three VGOS antennas (WESTFORD, KOKEE12M, and the 12 m antenna at the Goddard Geophysical and Astronomical Observatory, or GGAO12M) were added to a VLBI session (RD1810) that had six participating legacy S/X antennas to demonstrate the operational status of mixed-mode (VGOS plus legacy) observing and data processing. The data recording setup was the same as described above. However, the schedule was created to implement the original purpose of the session which was to observe sources that are weaker than usual for a geodetic VLBI session. As a consequence, the average signal-to-noise ratio (SNR) was lower than for the four KT sessions, yielding larger delay uncertainties. In addition, the longer scans (needed for the weaker sources), combined with longer time between observations (due to the legacy antennas being slower), resulted in fewer scans per hour, which adversely affected the geodetic results. The dates and some details of the five sessions are listed in Table 2.

**Table 2** Number of observations and sources for the KOKEE-KOKEE12M baseline of the KT and RD1810 sessions. The name of the database (first column) also encodes the date of the start of the session

| database name | session name | duration (hr) | number of observations | number of sources |
|---|---|---|---|---|
| 16MAR11VB | KT6071 | 0.75 | 17 | 16 |
| 16MAR18VB | KT6078 | 6.0 | 99 | 30 |
| 16MAR24VB | KT6084 | 4.4 | 81 | 30 |
| 16MAR30VB | KT6090 | 21.7 | 409 | 61 |
| 18DEC12XA | RD1810 | 24 | 41 | 17 |

The sequence of observations (the 'schedule') for all sessions was created using the standard geodetic VLBI program, *sked* (Gipson 2018). There are several antenna parameters that affect the number of observations and the distribution of the observations in azimuth and elevation, both of which are important for the robustness of the geodetic results. These are sensitivity, slew rates in azimuth and elevation, and the sky 'mask', which specifies the areas of the sky that are visible to the antenna. Of particular importance for these measurements is the close proximity of the two antennas, which results in blockage of the sky by KOKEE as seen from KOKEE12M. An early version of the mask, based on visual measurements of geometric blockage, was in place for the four KT sessions and resulted in no common visibility in the quadrant of the sky to the northwest, which contained 25 % of the potential visibility. Full sky measurements of the system temperature of KOKEE12M subsequent to these observations demonstrated that the blockage was not so severe, and the RD1810 session took advantage of this to improve the sky coverage.

Petrachenko et al. (2009) showed that the precision of the geodetic results improves as the number density (scans per hour) of observations at an antenna increases. This density is affected by many factors, including the minimum practical SNR per band, the minimum practical scan length, the sensitivity of the antennas, the strength of the available radio sources, and the slew rates of the antennas. The other important factor is the sky coverage, i.e., the distribution in azimuth and elevation of the observations. It is difficult to quantify the tradeoff among the different factors that will improve the precision of the geodetic results (Schartner et al. 2020).

While the minimum SNR for a reliable detection for a single observation is approximately seven, for these sessions the minimum SNR was set to 20 to allow for possible differences in the source flux densities from the catalogue values and for unmodeled variation in antenna sensitivity due to orientation or to weather conditions. Even though the minimum SNR could be achieved for many of the sources when using a scan duration of only a few seconds, a minimum scan length of 30 sec was specified for these sessions.

Another factor contributing to the number density of observations is the time required after the slew from the previous observation before data recording can begin. This time includes antenna settle time, setting of the gains and checking the digitization levels, system temperature measurement, and synchronization time at the correlator. While some of these activities may overlap in time, others must be sequential.

The candidate list from which the sources to be observed were selected was the catalogue used for scheduling the twice-weekly operational IVS geodetic sessions. Since this group of sources was selected for having compact structure on intercontinental baselines, most were expected to be unresolved for the much shorter intra-KPGO baseline.

For each session a 24-hour schedule was generated. For the sessions in 2016 the average number of scans per hour was 20, which is slightly greater than the 17 scans per hour of a recent global S/X geodetic session but is less than half that of the two-station (GGAO12M and WESTFORD) prototype VGOS sessions (Niell et al. 2018). This is due primarily to the limited slew rates of KOKEE, which are less than 2 °/s in both axes.

An exception to the operational configuration was made for KOKEE in that the phase-calibration signal was not enabled. It was necessary at that time to not have the phasecal signals present for both systems because the cross-correlation product of the tones at the two antennas, being coherent with the common 5 MHz frequency reference, would dominate the much weaker astronomical signal, thus preventing its detection. However, it was known from the long history of KOKEE that the signal chain electronics are sufficiently stable that a single, manually-determined correction can be applied to the phases of each channel for the duration of a legacy session with no significant loss of sensitivity or delay precision, thus allowing omission of the phase calibration signal. (However, see Section 6 for the possibility of a resulting systematic position error.)

## 4 Correlation and post-correlation processing

Recording and correlating data from the two different VLBI systems is referred to as 'mixed mode' since the radio signal is recorded in such different ways and requires significantly different correlator setup than either the legacy or VGOS processing. In addition, for the present results the signals are of two different polarization types, single right-circular for legacy and dual-linear for VGOS. At the time of correlating the tie data described here, it was necessary to make three separate passes to correlate the different combinations of recording types: legacy-legacy, VGOS-VGOS, and the mixed-mode legacy-VGOS. Improvements to the correlator program (DiFX), to the post-correlation program *difx2mark4*, and to the observable-estimation program *fourfit* (Cappallo 2017) have since been made which now allows all three combinations to be correlated in one operational pass.

The recorded modules for the two antennas were shipped to MIT Haystack Observatory, and the observations were correlated on a DiFX software correlator (Deller et al. 2011). The native output of the correlator (the so-called Swinburne files) were converted to Mark 4 format using the program *difx2mark4* in order to be compatible with the Haystack Observatory Processing System (HOPS) suite of programs (https://www.haystack.mit.edu/haystack-observatory-postprocessing-system-hops/). *difx2mark4* does more than just the format conversion, including normalization of the cross-correlation amplitudes using the auto-correlation for each station and correction for one- and two-bit sampling. The output of this program is operated on by the HOPS program *fourfit* to estimate, for each observation, group and phase delays, phase, delay rate, cross-correlation amplitude, and, for observations between VGOS broadband systems, a phase dispersion constant (more commonly referred to as the ionosphere delay). For a thorough description of the correlation and post-correlation procedures for VGOS processing, see Barrett et al. (2019).

The main HOPS tool is *fourfit*, and the most relevant options for these sessions include the application of phase calibration and the combination of the cross-polarization products. Phase calibration for KOKEE12M was applied in *multitone* mode which utilizes all tones within a channel to determine the single-station instrumental delay for that channel and a phase at the center of the channel. The baseline-differenced channel phases allow data from all channels to be combined coherently.

While the hardware phasecal system was not turned on for KOKEE, it was still necessary to provide values of the phase offsets of the channels in the form of fixed values. These were determined by using the observation of a source with high SNR to set the manual phasecal value for each channel as needed to reduce the fringe phase residuals to the *fourfit* model to zero.

For the sessions in 2016, keeping the phasecal at KOKEE turned off avoided having the astronomical cross-correlation signal dominated by the cross-correlated phasecal tones. For the session in 2018, RD1810, even though the phasecal hardware was not activated for KOKEE, spurious signals still appeared at multiples of 5 MHz that were strong enough to make the delays initially unusable. These were removed using the notch filter feature of *fourfit*, which resulted in a loss of approximately 5 % in SNR.

The program *fourfit* was used to coherently combine the XR and YR cross-correlation coefficients, where X and Y are the linear polarization data from the VGOS antenna and R indicates the sense of circular polarization of the data from the legacy antenna. (X, the sense of linear polarization, is not to be confused with X-band, the frequency range of the received radio signal.) Since the data are for X-band only, no dispersion (ionosphere delay) was estimated. The delay difference between the X and Y polarizations of the KOKEE12M antenna is only 0.14 ns, as determined from the correlations with the Westford data, as well as from previous and subsequent VGOS sessions. The Y minus X phase difference is approximately -50 degrees, also as determined by measurements over the history of the antenna. Both of these are corrected for in the *fourfit* processing.

After all phase corrections were evaluated, *fourfit* was run for all observations. A database in vgosDb format was then produced in three steps: vgosDbMake, vgosDbCalc, and vgosDbProcLogs (Bolotin et al. 2016). The first produces the skeleton database for all scans, the second adds the apriori values and partial derivatives, and the third adds the cablecal delays and meteorological information. As noted above, the cable delay measurements from the hardware cablecal system were applied for KOKEE, and the proxy cablecal delays were applied for KOKEE12M.

## 5 Geodetic analysis

In this paper geometric measurements, such as distances and physical cable lengths, are given in length units (e.g., millimeters), while time measurements, such as delays, are given in time units (e.g., picoseconds). For those two examples the conversion *in vacuo* is 1 mm = 3 ps.

The geodetic analysis was made with the program *nuSolve* (Bolotin et al. 2014 2019). This program operates on the vgosDb database to perform a least-squares estimation of many geodetic, geophysical, astronomical, and instrumental parameters using the square-root information filter (SRIF) algorithm. *nuSolve* can process only one session at a time, so each of the four days was processed separately.

The two KPGO antennas are separated by only 31 m, and the observations were made at centimeter wavelengths, so that *a priori* values for most astronomical and geophysical quantities were more accurate than could be estimated from these sessions. Therefore, the only quantities that were estimated were the geocentric coordinates of KOKEE12M relative to KOKEE and the variations in the clock and atmosphere delay differences between the antenna systems. The same position was used for KOKEE for all sessions, and the clock and atmosphere parameters for KOKEE12M were modeled as continuous piece-wise linear (PWL) functions with incremental rates. For such a PWL model the estimated values are, for each parameter, an initial value and rate for the first interval and a new rate for each of the successive equal-duration intervals. A fundamental question in the analysis is what intervals should be used for the clock and atmosphere delay differences. For the clock difference the electrical properties of the system components should be a guide, and for the atmosphere delay the physics of the atmosphere should be considered. In both cases the precision of the observations and the frequency of measurement should be taken into account.

Using a PWL function is only an approximation to the real physical changes. Since the clock and atmosphere delays are continuously variable, the shorter the interval is, the better the correspondence to the actual process. On the other hand, there should be a valid statistical basis for the number of parameters estimated; for example, in the absence of constraints among the parameters there cannot be more parameters than data.

### 5.1 Clocks

For the clock model, which describes the difference in time between the two systems, the main source of variability was the effect of changes in the ambient temperature on cables and components of the signal chain. These included the cable carrying the 5 MHz reference signal from the cablecal ground unit up to the phasecal generator for KOKEE12M and the cable bringing the signal coming down for KOKEE (see Figure 2 in Niell et al. 2018). This was verified for session KT6090 by comparison of the post-fit delay residuals of two solutions in which only a clock offset and rate difference were estimated but with the cablecal delays either used or not used. With no cablecal corrections, the post-fit delay residuals had a peak-to-peak range of about 200 ps with many changes in rate. Application of the cablecal corrections for the two antennas reduced the peak-to-peak range to about 80 ps, and the data could have been fit by only two rates. However, since the lengths of the two segments were unequal, but only equal-length segments can be modeled in nuSolve, a PWL interval shorter than a few hours was needed to avoid a large rate change within a segment. The phase delays are not as well corrected by the cablecal as are the group delays, perhaps due to thermally driven phase drifts in the LO signal in the KOKEE front-end. For the phase delays, a PWL clock interval of 30 minutes was necessary to absorb the remaining variations, and that interval was used for both the group and phase delay solutions of all sessions to provide consistency in the analysis.

## 5.2 Atmosphere delay

The delay of the radio waves by the neutral atmosphere is typically divided into the hydrostatic and wet components (e.g., Davis et al. 1985). Although the two antennas are separated by only 31 m, both components of the atmosphere contribute to the delay difference between the antennas.

The wet path delay as seen by a single antenna is highly variable, both in time and direction. It is modeled as a delay in the zenith direction (ZWD) times an elevation-dependent mapping function to give the line-of-sight delay at the time of each observation (e.g., Davis et al. 1985). The ZWD is parameterized as a PWL function of time.

The hydrostatic delay is assumed to be calculable from the pressure and is not estimated. The pressure is expected to have been measured for each observation with a barometer at each antenna. The only difference in pressure for two closely located antennas is due to their height difference because any horizontal pressure gradient is negligible over the 31 m separation. However, as there was only one set of meteorological instruments at KPGO, a correction for the expected pressure difference was needed.

With a scale height for the neutral atmosphere of approximately 8 km, and governed by hydrostatic equilibrium, the pressures were calculated to have a difference of 0.9 hPa as determined by the average pressure during the session KT6090 (887 hPa) and the vertical separation of the antennas (8 m). Thus, to correct for the height difference, the measured pressure values were adjusted by +0.9 hPa to obtain the pressure for KOKEE12M. With this pressure correction the mean estimated zenith atmosphere delay was 0.0 ps with a standard deviation of 1.1 ps, and the difference between making and not making this correction was 6.7 ps, which agrees within 0.1 ps with that expected from the height difference. The estimated vector position difference between using the uncorrected and the corrected pressures was less than 0.2 mm in all components.

The difference in the wet path delay seen by two antennas is a function of the separation of the antennas, the distribution of water vapor along the line of sight for each antenna, and the speed and direction of the wind carrying the water vapor (Treuhaft and Lanyi 1987). Since the atmosphere information (water vapor and wind) is not known, we have evaluated the PWL interval empirically by examining the weighted root-mean-square (WRMS) of the post-fit delay residuals of the geodetic solution, the additive noise required to achieve chi-squared per degree of freedom (chi2pdof) of approximately 1.0, and the change in parameter values as the length of the PWL interval was reduced. Intervals of 60, 30, 20, and 15 minutes were used for KT6090, keeping the PWL interval for the clock at 30 minutes. Table 3 shows the resulting WRMS post-fit delay residuals and the additive delay noises.

**Table 3** Atmosphere delay PWL interval length evaluation. The WRMS and the additive delay noise required to achieve chi2pdof of approximately 1.0 for the phase delays for different PWL intervals of the ZWD (session KT6090)

| ZWD interval (min) | WRMS(ps) | additive delay noise (ps) |
|---|---|---|
| 60 | 4.1 | 4.5 |
| 30 | 3.9 | 4.4 |
| 20 | 3.6 | 4.2 |
| 15 | 3.3 | 3.9 |

For these four solutions the estimated components of the baseline varied by less than 0.3 mm, and the uncertainties varied by less than 0.02 mm. Thus, from the geodetic point of view, at the level of 0.1 mm in an RMS sense for the component values, the exact interval is not significant.

On the other hand, from the point of view of understanding the magnitude of the contribution of the error sources, the reduction of the additive noise might be interpreted as demonstrating that the shorter PWL intervals are a better indication of the variability time scale of the atmosphere. However, it is possible that the shorter intervals are also absorbing some of the variation of the clock difference that is not accounted for by the 30 minute PWL interval used for the clock parameter. The results reported in the remainder of this paper are for a PWL interval of one hour for the ZWD and 30 minutes for the clock. For RD1810 there were insufficient observations for so many parameters, so the PWL interval was set to six hours for the clock and 24 hours for ZWD.

In view of the small distance between the antennas, a spatial gradient in the atmosphere delay, for example as modeled by Chen and Herring (1997), was not considered.

## 5.3 Relative merit of group and phase delay

Both the group delay and phase delay X-band observables are provided in the correlator output and can be used for the geodetic analysis. However, the uncertainty (sometimes referred to as formal error) of the phase delay calculated by *fourfit* is smaller than that of the group delay by a factor of approximately 1/15. The median group and phase delay uncertainties for KT6090 are 12 ps and 0.8 ps, respectively. Ninety-seven percent of the phase delay uncertainties are less than 2 ps. Thus, it is preferable to use the phase delays for the geodetic solution. The benefit can be seen in Fig. 2 which shows histograms of the uncertainties (left) and post-fit residuals for the geodetic estimation (right) for the group and phase delays for KT6090.

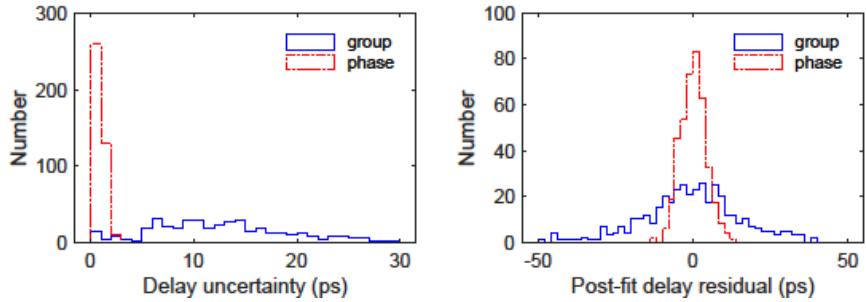

**Fig. 2** Histograms of (left) phase-delay and group-delay uncertainties (*fourfit* output) and (right) re-weighted phase- and group-delay post-fit residuals, all for KT6090

The drawback of phase delays vs. group delays is that the former are ambiguous by an integer number of cycles of phase at the reference frequency. In general, the challenge to using phase delays is the lack of sufficiently accurate *a priori* values for the model parameters, including clock and atmosphere delays, to determine residuals to the model of less than half an ambiguity spacing (half of one cycle of phase). This spacing is only 122 ps, corresponding to about 36 mm for the geometric model. As mentioned above, the close proximity of the antennas means that the accuracies of the catalogues of source positions and geophysical parameters, which are obtained from observations that span intercontinental distances, provide good *a priori* values, and in practice they are sufficiently accurate to not introduce ambiguities. For the atmosphere delay the proximity of the antennas means that the hydrostatic delay differences are less than 10 ps, as explained above, which is also much less than half an ambiguity. The wet delay difference is even less.

### 5.4 The estimation procedure

Obtaining the final geodetic estimate utilized multiple steps, proceeding from single-band delay to multiband delay to phase delay, i.e. from least precise to most precise with each step refining the clock and atmosphere models. (See Herring (1992) for a description of phase delay results on a short baseline.) Both single-band and multiband delay are a type of group delay, which is the derivative of phase with frequency. Single-band delay is the average over all channels of the group delay over the 8 MHz span of the individual channels, while multiband delay is the group delay of channel phase versus RF frequency. For our tie observations, the delay uncertainty (standard error) reported by *fourfit* is $\sqrt{12}/(2\pi*SNR*8MHz)$ for single-band delay, $1/(2\pi*SNR*RMS\_frequency)$ for multiband delay, and $1/(\pi*SNR*ref\_frequency)$ for phase delay, where $8\,MHz/\sqrt{12}$ is the RMS of the frequencies of a single channel, RMS_frequency is the RMS of the X-band channel frequencies (280 MHz), and ref_frequency was set to the lower edge of the lowest frequency channel (8212.99 MHz).

The single-band delays are unambiguous, so those values were used to obtain an initial quadratic clock model for KOKEE12M (KOKEE was set as the reference station). Those parameters are used as the *a priori* clock model for the group delays to clarify, if needed, the identification of any group delay ambiguities. The group-delay ambiguity spacing for the X-band channel distribution is 50 ns, but no ambiguities were detected for any of the sessions because the *a priori* values for all models, with the inclusion of the clock model from the single-band delays, were much better than this. Next the PWL interval values to be used for the clock and atmosphere delay were set, and those parameters and the position of KOKEE12M were estimated. An initial solution was obtained, generally resulting in a very large value for the reduced chi-squared of the post-fit delay residuals. This was a consequence of the delay uncertainties produced by *fourfit* being based only on the SNR calculated from the number of bits recorded, while effects such as atmosphere and clock variations cause differences to the model that are much larger than those SNR-based uncertainties. To determine more realistic uncertainties that account for all sources of error, an iterative procedure was followed in which, at each step, increasing amounts of delay noise were added in quadrature to the SNR-based uncertainties, a new solution was run, and chi2pdof was re-calculated. This cycle was continued until chi2pdof was approximately 1.0. Any post-fit delay residuals greater than 3.5 times their re-weighted uncertainty were marked for exclusion. The estimation/re-weight/outlier-check sequence was repeated until no outliers were detected, yielding the final solution.

Analysis of the phase delays differed from the group delay analysis only in that there were many points with different ambiguity assignments due to the smaller spacing of the ambiguities. Application of the automatic ambiguity resolver in *nuSolve* was not successful, so the residual phase-delay ambiguities that were present required manual inspection and correction. Following that adjustment, the procedures were the same as for the group delay estimation. The WRMS of the post-fit delay residuals and the additive delay noise required to achieve chi-squared per degree of freedom near 1.0 are given in Table 4.

================================================================================

**Table 4** WRMS post-fit delay residuals (pfdr) and additive noise for each session for group and phase delay. For KT sessions the PWL interval for the clocks was 30 minutes and for the zenith atmosphere delay was 60 minutes. For RD1810 the intervals were 6 hours and 24 hours, respectively. The WRMS and additive noise are larger for 16MAR24 due to bad weather. The units are picoseconds.

| Session | Group delay | | Phase delay | |
|---|---|---|---|---|
| | WRMS pfdr | additive noise | WRMS pfdr | additive noise |
| 16MAR11 (KT6071) | 13.8 | 14.6 | 2.4 | 3.4 |
| 16MAR18 (KT6078) | 16.6 | 14.7 | 3.3 | 3.7 |
| 16MAR24 (KT6084) | 24.2 | 18.4 | 4.4 | 4.9 |
| 16MAR30 (KT6090) | 14.0 | 10.6 | 4.2 | 4.5 |
| 18DEC12 (RD1810) | 34.5 | 29.3 | 6.2 | 7.0 |

================================================================================

The resulting geocentric and topocentric coordinates and the corresponding uncertainties as determined from the phase delays are given in Table 5 for each session. For comparison, the coordinate uncertainties for the group delay solution for the fourth session, KT6090, are larger by a factor of 3.6 in all components.

The group delay coordinate values are consistent with the phase delay results. None of the East or North components or the lengths differ from the weighted mean phase delay values by more than the group delay uncertainty. For the Up component, only two of the five values differ by more than 1.4 times the group delay uncertainty.

**Table 5** Geocentric and topocentric coordinate estimates from phase delay. In the top half are the geocentric components and lengths of the baseline obtained in the nuSolve phase delay solutions for the baseline vector from KOKEE to KOKEE12M. The entries in the lower half are obtained by a rotation to the local topocentric frame. The full covariance in XYZ was transformed to derive the topocentric component uncertainties. No corrections have been applied. The units are millimeters.

| Session | X | $\sigma_X$ | Y | $\sigma_Y$ | Z | $\sigma_Z$ | L | $\sigma_L$ |
|---|---|---|---|---|---|---|---|---|
| 16MAR11 (KT6071) | 6069.92 | 2.49 | -19215.51 | 1.04 | -23719.92 | 1.26 | 31124.18 | 1.02 |
| 16MAR18 (KT6078) | 6073.45 | 1.12 | -19212.79 | 0.49 | -23721.96 | 0.46 | 31124.74 | 0.38 |
| 16MAR24 (KT6084) | 6069.09 | 1.47 | -19216.10 | 0.66 | -23720.86 | 0.70 | 31125.10 | 0.57 |
| 16MAR30 (KT6090) | 6071.91 | 0.63 | -19214.15 | 0.28 | -23721.64 | 0.27 | 31125.04 | 0.22 |
| 18DEC12 (RD1810) | 6071.68 | 4.25 | -19214.02 | 1.83 | -23721.97 | 1.66 | 31125.16 | 1.19 |

| Session | E | $\sigma_E$ | N | $\sigma_N$ | U | $\sigma_U$ | L | $\sigma_L$ |
|---|---|---|---|---|---|---|---|---|
| 16MAR11 (KT6071) | 20127.29 | 0.54 | -22344.37 | 0.52 | -8020.94 | 2.88 | 31124.18 | 1.02 |
| 16MAR18 (KT6078) | 20125.97 | 0.25 | -22344.66 | 0.29 | -8025.65 | 1.25 | 31124.74 | 0.38 |
| 16MAR24 (KT6084) | 20127.55 | 0.38 | -22345.62 | 0.34 | -8020.39 | 1.68 | 31125.10 | 0.57 |
| 16MAR30 (KT6090) | 20126.71 | 0.14 | -22345.09 | 0.15 | -8023.76 | 0.71 | 31125.04 | 0.22 |
| 18DEC12 (RD1810) | 20126.50 | 0.74 | -22345.46 | 0.62 | -8023.72 | 4.82 | 31125.16 | 1.19 |

It is of interest to evaluate the benefit of increasing the session duration for reducing the uncertainties. While it is clear that uncertainties are inversely related to the duration of the session (not including RD1810), it is reasonable to ask if the uncertainty is decreasing as expected. Comparing only the three longest KT sessions, the answer is affirmative: the uncertainties scale inversely as the square root of duration to better than 20 % for KT6078 and KT6084 compared to KT6090.

From the phase delays it can be concluded that an unmodeled delay noise of only approximately 4 ps would explain the scatter in the delays on a time scale of less than 30 minutes. Since the group delays require a much larger unmodeled delay error to achieve a chi2pdof near one for the same clock and ZWD parameterization, one or more additional unmodeled noise sources are required to explain the group delay scatter. The phase delays allow an upper limit of 4 ps to be set on an atmospheric or clock origin for the noise, so for the group delays, the extra noise must be due to the instrumentation.

The Treuhaft and Lanyi (1987) atmosphere wet delay model predicts a variation of a few picoseconds over a few minutes, which is the scan-to-scan time for these sessions, so the unmodeled phase delay noise might reasonably be attributed to the simplified treatment of the atmosphere delay (using PWL instead of a continuous stochastic model). Of course, some part of the observed variation is likely to be due to instrumentation since the clocks are also modeled as piece-wise linear.

## 5.5 Other models to account for the unexplained additional phase-delay noise

No correlation that is dependent on the time or angular separation between the observations was incorporated in the estimation process in *nuSolve*. As an alternative to, and more realistic than, re-weighting by adding delay noise to the observed uncertainty for each observation as was done for the present analysis, correlations based on a turbulence model for the atmosphere delay could be used in the estimation process. Halsig et al. (2016) demonstrated that such correlations, whose source is attributed primarily to the structure of the atmosphere and for which the application to VLBI was quantified by Treuhaft and Lanyi (1987), can achieve a value of one for chi2pdof for the post-fit delay residuals within a session without the addition of delay noise. In addition, the scatter in baseline lengths among the set of antennas that was considered was found to have a chi2pdof of close to one when calculated using the session uncertainties.

Gipson (2007) showed that the incorporation of station-dependent clock and atmosphere delay noises and their correlation among stations in a scan gives more realistic formal errors and reduces the scatter in baseline length without the addition of a separate arbitrary additive noise on each baseline. Halsig et al. (2016) concluded that the Gipson (2007) model provided comparable results to the turbulent atmosphere model while being much simpler to implement. For a scan involving only two antennas this model would reduce to the addition of a constant noise component, which is equivalent to the additive noise described above, and an elevation-dependent component, which is not included. Given the high precision of the phase delays, it will be interesting to investigate the inclusion of the elevation-dependent noise term in future observations.

Halsig et al. (2019) utilized the method devised by Halsig et al. (2016) to study the separation of instrumentation and atmosphere as the source of noise on the short baseline between two antennas at the Wettzell Observatory, incorporating a third, distant antenna at the Onsala Space Observatory to provide a more absolute basis for the atmosphere components. From both the short 123 m baseline intra-Wettzell observations, and by differencing the long baseline observations, they arrived at a tropospheric noise on the short baseline of 3-10 ps, with which the results reported here (Table 3) are entirely consistent.

A more realistic treatment of the noise model is not expected to change the results at more than the level of the uncertainties, so the primary benefit of introducing a more accurate noise model would be to improve our understanding of the sources of the unexplained noise.

## 6 Non-stochastic errors

We discuss in this section the effect of errors due to delay variations in the frequency- and signal-carrying cables and to mechanical and thermal deformation of the antennas.

### 6.1 Un-corrected delay variations

In addition to the noise-like errors of the observations, there are possible systematic errors that are not accounted for and that deserve further investigation. A primary candidate is an un-modeled delay that repeats independent of the time of the observation but varies with antenna orientation. This was found to be a source of significant error in the case of the 5 MHz reference frequency cable at GGAO, which has an azimuth- and elevation-dependent delay that is mitigated by using a proxy cable delay correction (Niell et al. 2018). For the Kokee tie sessions, an example is delay variations in the cable bringing the signal from the KOKEE receiver to the control room. Because the KOKEE phasecal was turned off, phasecal phases could not be used to correct for these cable variations. Instead, the KOKEE cablecal delays were applied to the group and phase delays (see section 5.1) to correct for the receiver-to-control-room cable variations on the assumption the delay variations in the reference frequency cable, as measured by the cablecal, were similar to those in the receiver-to-control-room cable. That assumption appears to have been reasonably well founded for the slow, temperature-induced drifts in group delay (section 5.1). However, the assumption may not have been valid for the potentially more serious orientation-dependent delay variations, as described in the next paragraph.

An investigation of the orientation dependence of the KOKEE phasecal delay and cablecal measurements at X-band for several S/X sessions revealed that only the former varied by more than a few picoseconds, and then only in elevation angle (Corey 2018). Taken together, these facts indicate the presence of elevation-angle dependence in the signal chain delay somewhere in the receiver or in the cable from the receiver to the control room. For the best determined session, the variation in phase delay was consistent with a sin(elevation) term of amplitude 6±1 ps. Were this effect to apply to the KT sessions (and there is no reason to exclude that possibility since the KOKEE system was not modified between the time of the KT sessions and the S/X sessions in the Corey (2018) study), the estimated height of KOKEE in local coordinates would be decreased, and the Up component of the baseline would be changed by approximately +2 mm. Since this effect cannot be verified as applying to the KT sessions, no correction has been made to the Up component of the baseline reported here, but it must be considered when assessing the uncertainty and the agreement with other measurements, such as an optical survey.

The other part of the signal path that is not calibrated or corrected is from the feed to the injection point of the phasecal system, a total distance of only tens of centimeters. The feed, cables, and connectors are mechanically stable and not subjected to stress, so there should not be any antenna-orientation dependence. The difference in path length between the two polarizations for KOKEE12M has been calculated from the data for every session since its installation, and both the delay and phase differences are stable at the level of less than a few picoseconds of delay and a few degrees of phase, thus providing some evidence for the lack of variation.

Another possible source of antenna-orientation-dependent delay variation is the mixing of circular polarization on KOKEE and dual-linear polarizations on KOKEE12M. However, since the antennas are only 31 m apart, the relative parallactic angles do not vary by more than 0.01 degrees through a session. Thus, any effect of differential feed rotation on the phase delays is in the femtosecond range.

## 6.2 KOKEE12M gravitational deformation

A common mechanical deformation is gravitational distortion of the antenna. This usually varies primarily with elevation for an azimuth-elevation-type antenna (e.g., Carter et al. 1980; Sarti et al. 2011; Lösler et al. 2019) and manifests itself as a change with elevation of the path length through the antenna optics to the feed. If undetected, this change leads to an error in the estimated height of the intersection of axes relative to a fixed point on the ground. This error can only be determined by external measurements of the change in shape of the antenna. Lösler et al. (2019) measured the change in the shape of one of the 13.2 m VGOS antennas at the Onsala Space Observatory (OSO), Sweden, and found that the change in delay from 5° to 90° elevation angles is less than 1 mm, which would result in a vertical height error of less than 1 mm. Finite element modeling by the manufacturer of the new 12 m VGOS antenna at McDonald Observatory, Texas (Merkowitz et al. 2018), which is identical to KOKEE12M, indicated that it should also have a sub-millimeter change in delay over the observing range for geodetic observations.

## 6.3 KOKEE gravitational deformation

The principal contributions to the deformation effects are the change in the shape of the main reflector, which flattens as the antenna moves from 0° to 90° elevation, and the change in the distance from the vertex of the antenna to the subreflector or prime-focus feed, which decreases over that range. There is not yet either a model for, or measurements of, gravitational deformation of the KOKEE antenna. The only comparable antenna for which measurements exist is the 20 m antenna at OSO.

From an extensive study making use of detailed measurements of the antenna structure, and basing evaluation of the path length changes on variations in raypath centroid rather than focal length, Bergstrand et al. (2019) estimated that the path length traversed by the radio signal through the antenna decreased by 9±2 mm over the elevation range 0° to 90°. KOKEE differs from the Onsala antenna in being a prime focus rather than Cassegrain system. Thus, the vertex-to-subreflector contribution to the path length change would enter only as the change in one-way distance, which for the Onsala antenna is approximately 2 mm. Although the value of the same effect for KOKEE depends on the actual structure, if KOKEE is assumed to be structurally similar to the Onsala antenna, an estimate of the total path length change for KOKEE would be on the order of -7 mm.

A different estimate of the effect of gravitational deformation on KOKEE can be obtained from measurements of two larger (32 m) prime focus antennas at Medicina and Noto, Italy, for which the path length changes were found to be +10 mm and +7 mm, respectively (Sarti et al. 2011). However, using scaling based on structural mechanics, they derived an approximation for antennas that are of the same design but different diameter that gives the deformation as being proportional to the square of the diameter. Applying this scaling would give a path length change of between +3 mm and +4 mm for KOKEE.

A possible explanation for the difference in sign of the two estimates may be found in the warning in Bergstrand et al (2019) of the importance of using the raypath centroid to calculate the path delay changes rather than, as done by Sarti et al (2011), using geometric optics based on focal length change as formulated by Clark and Thomsen (1988). Regardless of which is more correct, it is clear that there is an uncertainty of up to a centimeter in the path delay changes due to gravitational deformation for KOKEE. For both cases above, the functional form of the path length change can be modeled with a sin(elevation angle) dependence, resulting in an estimated height change closely equal to the difference in path length at 90° elevation compared to 0°. Thus, a conservative estimate for the uncertainty in the effect of gravitational deformation can be taken as 10 mm in the local vertical. Regardless of the details of the calculations above, it is the large uncertainty of this one effect compared to the VLBI uncertainties that is a cause for concern.

## 6.4 Thermal deformation

The antenna structure and the supporting foundation will also experience deformation due to changes in ambient temperature. Nothnagel (2009) describes the standard IVS models to be used for calculating the effect of the deformation on the delay for the different types of antenna mounts, including the azimuth-elevation type of the two VLBI antennas at KPGO. It would be most accurate to implement the correction as an adjustment to the delay for each observation. However, as shown below, the total effect is at the level of only a few tenths of a millimeter, so it is unlikely that correcting at the observation level would result in a significant difference to an average correction over a session. Instead the thermal deformation has been implemented as a height change model by evaluating the deformation delay at zenith and multiplying by the speed of light.

An important part of the standard model is the requirement to refer all measurements to a common reference temperature so that measurements made at different temperatures can be compared and combined. The reference temperatures for antennas used in the analysis of IVS observations, along with the recommended structural parameters, are contained in the file *antenna-info.txt* (https://raw.githubusercontent.com/anothnagel/antenna-info/master/antenna-info.txt). These parameters include the dimensions and coefficients of expansion with temperature for the major structural components, as well as a reference temperature. The reference temperature is obtained from the Global Pressure and Temperature (GPT) model (Böhm et al. 2007) using the recommended modified Julian day (MJD) epoch of 44357.3125 (Wresnik et al. 2007).

For each of the sessions the mean temperature was used for the correction to the reference temperature. To allow for thermal lag in the antenna structure, the model given by Nothnagel (2009) includes latency values, i.e., a time delay, but these were not utilized because the temperatures preceding the sessions were not available. Neglecting latency would have a larger effect the shorter the sessions. For example, for the KT6071 session (Table 2), which had a duration of only 45 minutes, the temperatures that affected the observations occurred several hours prior to the session and could have differed from the mean temperature by as much as 4 °C, resulting in a height correction up to 0.4 mm. On the other hand, for KT6090, which lasted 22 hours, the difference between actual and latent temperature would average to much less than 1 °C over the session.

For the KOKEE position given in the configuration file (https://ivscc.gsfc.nasa.gov/stations/config/ns/kokee.config.txt), the GPT reference temperature is 17.2 °C. However, the reference temperature for KOKEE that was entered in `antenna-info.txt` sometime in the past and is in common use is 16.9 °C. Since the difference in height adjustment between these two values is much less than 0.1 mm, the value of 16.9 °C is used in this paper in order to retain consistency with other analyses in which the height adjustments for KOKEE have been made using `antenna-info.txt`. Because of the sensitivity to the reference temperature, it is imperative that the values for all stations in any IVS geodetic session be explicitly documented in the reported analysis.

The mean temperatures varied among the KT sessions by only 3 °C, and the largest difference to the reference temperature was +4 °C. The height adjustments for KOKEE ranged from -0.4 mm to -0.1 mm, while for KOKEE12M all corrections were less than 0.1 mm. The total KOKEE12M-KOKEE height adjustment ranged from 0.1 mm to 0.3 mm. Since the reference temperature is higher than the temperatures at the time of observation, it might seem counter-intuitive that the height adjustment has the effect of decreasing the height of the antennas. As pointed out by Wresnik et al. (2007), however, the effect of a temperature increase on the feed support struts acts to change the height correction in the opposite direction to the expansion of the pedestal and other structural elements supporting the main reflector surface, which for these two antennas more than compensates for the primary structural expansion. (It is important to notice that, in the initial publication of the thermal correction formula (Nothnagel et al. 1995), the sign of the strut correction term was incorrect. Therefore, for completeness we reproduce in the Appendix the deformation model as applied to these observations.)

## 6.5  Antenna tilt

The position of an antenna will change temporally relative to a local frame if the antenna pedestal tilts with time, due, for example, to differential compaction of the soil under the antenna foundation. Therefore, position determinations made at different times should be corrected to a common epoch for accurate comparison. This time-dependent change is in contrast to a constant error due to neglect of the gravitational deformation of an antenna. Apparent local motion such as this is most directly and accurately measured by optical means using nearby reference markers (e.g., Erickson and Breidenbach 2019). However, dedicated surveys are time-consuming and therefore are not repeated frequently. An alternative approach is the use of a permanently installed Vector Tie System that continuously measures the position of the intersection of axes relative to a local network of markers, thus providing unambiguous, three-dimensional information. Lacking such measurements on a regular basis, local horizontal motion due to varying tilt of the antenna structure can, in principle, be detected by monitoring the antenna pointing errors in azimuth and elevation. These reflect deviations of the antenna structure from an ideal model, including such things as non-orthogonality of the axes. A change in the tilt of an antenna has a specific signature in the pointing errors that develop as a consequence.

Pointing measurements have been used to investigate possible tilt change for the two KPGO antennas. For the KOKEE data, which span eighteen years from 2001 to late 2019, a model of linear motion at the height of the intersection of axes (the antenna reference point) yields a rate of 0.01±0.03 mm/yr in both coordinates. Thus, for the two years between the mean epochs of the VLBI and NGS measurements of the KOKEE to 12 m vector, any tilt change of the KOKEE antenna would have contributed less than 0.1 mm to the uncertainty in the difference in the baseline vector as measured by the two methods.

The picture of the data for KOKEE12M for the four years following its installation is less clear. The eastward motion is consistent with a rate of 0.30±0.08 mm/yr with chi2pdof of 1.0, although a step-function change of approximately 1 mm in 2017 or 2018 cannot be ruled out. The pointing corrections for the north direction over the 3.5 years from late 2016, which is after the KT sessions, through 2020, are well fit by a linear motion having a rate of -0.11±0.02 mm/yr. However, measurements in early 2016, just prior to the KT sessions, and in mid-2016, when combined with those of late 2016 are clearly at odds with this model, having a rate of +2.1 mm/yr that is equally consistent with linear motion over that interval. If the tilt measurements near the mean epochs are assumed valid, the north movement over the two years between the KT sessions and the NGS2018 value could be as large as approximately 1.1 mm.

Lacking any corroborating measurements for either the east or north tilt variation of the 12 m antenna, the rates have been assumed to be zero, and the uncertainties of the east and north component differences have been enhanced by the addition of 1 mm in quadrature to the other uncertainties.

# 7 Geodetic results

## 7.1 Mean topocentric position

The values of chi2pdof across all sessions for the components of the geocentric baseline vector, relative to their weighted means, are 1.5, 4.3, and 0.9 for the X, Y, and Z components, respectively (Table 5). The corresponding values for the East, North, and Up topocentric components are 3.7, 1.7, and 1.8. These values indicate that, under the assumption that the baseline has not changed over the span of these sessions, the uncertainties are underestimated for most components. The method of including arbitrary unknown delay-like noise that has an effect on an intra-session time scale (one minute to 12 hours) does not necessarily account for errors on the inter-session time scale (more than one day). In a manner similar to the quadrature addition of a fixed amount of delay noise to the delay uncertainty for each observation to achieve chi2pdof near a value of one for the post-fit delay residuals, a measurement 'noise' can be added in quadrature to all session measurements of each component. For an added noise of 0.5 mm the resulting values of chi2pdof for the components relative to the weighted mean values for East, North, and Up are in the range 0.6 to 1.6, where the weighting is composed of the combined per-session uncertainty and added noise.

The residuals of the topocentric phase delay coordinate estimates (including the adjustment to the reference temperature for the thermal deformation) are shown in Fig. 3 for each session relative to the re-weighted (i.e., including additive noise) means over all five sessions. The adopted uncertainties in the weighted mean components were obtained by multiplying the uncertainties from the weighted mean calculation by the square root of the chi-squared per degree of freedom. These uncertainties are insensitive to the amount of added noise, changing by less than 0.1 mm for a factor of two change in the added noise.

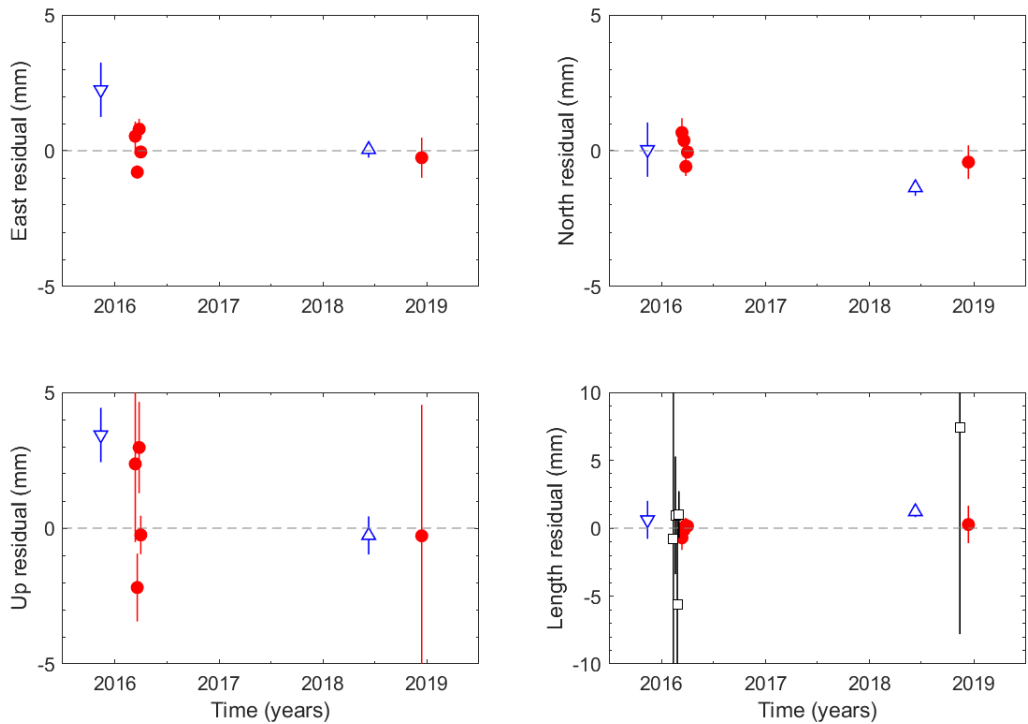

**Fig. 3** Residuals to the weighted mean phase delay solution for the components and length of the topocentric vector from KOKEE to KOKEE12M: Symbols: red circles – phase delay; black squares – group delay (slightly offset in time for clarity); blue triangles – independent measurements from two optical surveys: downward pointing for 2015; upward pointing for 2018. Frames: top left – East; top right – North; bottom left – Up; bottom right – Length (note different vertical scale)

## 7.2 Total topocentric uncertainties

To arrive at an estimate of the accuracy of the tie vector between the two antennas, the systematic errors described in Section 5 must be included. The result is summarized in Table 6.

Table 6 Summary of VLBI uncertainty estimates in the topocentric frame (mm)

|  | East | North | Up |
|---|---|---|---|
| VLBI phase delay weighted mean | 0.3 | 0.2 | 0.8 |
| KOKEE instrumental delay orientation dependence | 0.0 | 0.0 | 2 |
| KOKEE gravitational deformation | 0.0 | 0.0 | ~10 |
| Combined thermal deformation | 0.0 | 0.0 | 0.1 |
| KOKEE12M tilt | 1.0 | 1.0 | 0 |
| TOTAL | 1.0 | 1.0 | ~10 |

It is important to note that, although the precision of the VLBI tie is under 1 millimeter, the limitation for use in construction of the TRF is the possibility of systematic errors, such as gravitational deformation and the instrumental delay orientation dependence. Even accounting for these, the total uncertainty in the topocentric positions is less than 2 mm in the horizontal and only about 10 mm in the local Up direction. Note that there is the potential for reducing the error in the vertical component by almost an order of magnitude if a survey were conducted to determine the gravitational deformation of the KOKEE antenna.

## 8  Comparison of VLBI and optical survey vector ties

Intra-site ties among geodetic instruments of different types (GNSS, VLBI, SLR, and DORIS) are made by optical surveys since, as yet, there is no measurement technique that allows the relative positions of the diverse instrumentation to be measured in the native observing mode. Even for intra-technique, e.g. GNSS-GNSS or VLBI-VLBI, the optical surveys are generally as, or more, accurate than the native observations. At the very least they provide an independent measurement. Thus, it was planned for an optical survey to be made of the KOKEE12M antenna relative to the other geodetic instruments when the antenna became operational.

### 8.1  Tie results from optical surveys

Optical surveys among the many geodetic instruments of the KPGO complex, including both KOKEE12M and KOKEE, were made in 2015 November 10-13 (Carpenter 2016) and in 2018 June 7-13 (Erickson and Breidenbach 2019), before and after replacement of the KOKEE azimuth-angle bearing, respectively. An earlier survey in 2014 (Fancher et al. 2015) did not include KOKEE12M as it had not been constructed, but it does provide a comparison for the possible height change due to the bearing replacement since for both the 2014 and 2018 surveys the GPS antenna designated KOKB was used as the common reference point.

In Table 13 of the report for the 2018 survey, the difference in the height of KOKEE relative to KOKB between the 2018 and 2014 surveys, in the sense 2018 minus 2014, is given as (-0.9, 0.5, 1.0) mm in (East, North, Up) when both are transformed to ITRF2008. For the 2014 survey the uncertainty in the Up component of KOKEE relative to KOKB was approximately 0.3 mm as calculated from the reported covariance matrix. For the 2018 survey the uncertainties were 0.3 mm for East and North and 0.7 mm for Up.

For the 2014 survey the height correction to the reference temperature for KOKEE was reported to be +0.2 mm by Fancher et al. (2015) (the temperature at the time of measurement was lower than the reference temperature), but it was not added to the survey results. No height correction was made for the 2018 survey, but we have calculated the correction using temperatures at the time of observation obtained from Erickson and Breidenbach (private communication); 0.6 mm should be subtracted from the reported values. Combining the two reduces the change in Up to 0.2 mm. The resulting estimate for the height change is then 0.2±0.8 mm.

Thus, since the uncertainties in the 2015 and 2018 surveys are comparable to the inferred height change, it is reasonable to average the measurements of the KOKEE-KOKEE12M position difference before and after the azimuth bearing change. Similarly, the results of the RD1810 measurement have been included in calculating the weighted mean position for the VLBI results.

## 8.2 Comparison of VLBI and the optical surveys for the KOKEE to KOKEE12M vector

The uncertainties for the position of KOKEE12M relative to KOKEE for the 2015 NGS survey were 1.0 mm for all components (Carpenter 2016). The covariance matrix provided for the 2018 NGS solution (Erickson and Breidenbach 2019) was used to obtain the uncertainties of the KOKEE-KOKEE12M baseline for that session since the positions of all of the instruments were reported relative to the KOKB GPS antenna. The 2018 vector was adjusted to a reference temperature of 16.9°C. No adjustment was made for the 2015 survey since the temperature was not included in the report. The NGS weighted mean is given in Table 7 as is the weighted mean of the VLBI phase-delay results corrected to the same reference temperature for each session. The uncertainties in the VLBI results that allow for possible systematic errors (as given in Table 6) are enclosed in parentheses. The differences of the weighted means, in the sense NGS minus VLBI, are given in the bottom row.

================================================================================

**Table 7** Topocentric (ENU) and length estimates of the KOKEE to KOKEE12m baseline vector from optical surveys and VLBI observations. The weighted mean date of the VLBI observations is 2016 April 11 20:34 UTC. (All units are millimeters.)

|  | E | $\sigma_E$ | N | $\sigma_N$ | U | $\sigma_U$ | L | $\sigma_L$ |
|---|---|---|---|---|---|---|---|---|
| NGS 2015 | 20129.0 | 1.0 | -22345.0 | 1.0 | -8020.2 | 1.0 | 31125.5 | 1.4 |
| NGS 2018 | 20126.8 | 0.3 | -22346.4 | 0.3 | -8024.0 | 0.7 | 31126.1 | 0.4 |
| wtd mean | 20127.0 | 0.3 | -22346.3 | 0.3 | -8022.8 | 0.6 | 31126.1 | 0.4 |
| VLBI | 20126.7 | 0.3(1.0) | -22345.1 | 0.2(1.0) | -8023.7 | 0.8(~10) | 31125.0 | 0.3(0.3) |
| NGS-VLBI | 0.2 | 0.4(1.0) | -1.3 | 0.4(1.0) | 0.8 | 0.8(~10) | 1.2 | 0.4(0.4) |

================================================================================

The weighted means of the optical surveys (row 3) and the VLBI measurements (row 4) differ by less than about three times the combined stochastic uncertainties in any component. For the Up component this is perhaps fortuitous given the possibility of correction for either or both of the KOKEE instrumental delay and gravitational deformation by up to approximately 10 mm.

To provide input to ITRF2020 the weighted mean topocentric VLBI vector and covariance were rotated to the geocentric frame. The results are given in Table 8.

**Table 8** Weighted mean geocentric vector (mm) from KOKEE to KOKEE12m and the off-diagonal elements of the covariance matrix (mm^2) as measured by VLBI.

|  | X | $\sigma_X$ | Y | $\sigma_Y$ | Z | $\sigma_Z$ | L | $\sigma_L$ |
|---|---|---|---|---|---|---|---|---|
| wtd mean | 6071.9 | 0.7 | -19214.1 | 0.4 | -23721.6 | 0.4 | 31125.0 | 0.3 |
|  | $\sigma_{XY}$ | | $\sigma_{XZ}$ | | $\sigma_{YZ}$ | | | |
| covariance | 0.1567 | | -0.1678 | | -0.0598 | | | |

================================================================================

## 9 Discussion and conclusions

It is important to link the VGOS broadband and legacy S/X networks as quickly as possible in order to incorporate the expected higher accuracy of the VGOS antennas into the ITRF. An important contribution to this tie is the measurement of the baseline vectors between co-located VGOS and legacy VLBI antennas. The results reported here demonstrate that such ties can be made with sub-millimeter precision by making only an easily-configurable change in the frequencies of the VGOS system and scheduling the co-located antennas to take advantage of their common full sky coverage. The use of X-band-only phase delays provides the reduced uncertainties (compared to group delays) that enable the high baseline vector precision. The alternative of adding the VGOS antennas to larger networks of legacy antennas observing in S/X mode, as for RD1810, cannot yield the same observation precision or temporal scan density due to the generally lower correlated flux densities on the longer baselines and to the lower delay precision of the ionosphere-corrected group delays. While that session offered the potential to measure any change in the position of the 20 m antenna resulting from the azimuth-bearing modification, the large uncertainty of almost 5 mm in the local vertical is much greater than the expected change of less than a few millimeters from the mechanical modification to the bearing surfaces. Thus, it is important to execute another set of tie measurements that will achieve or exceed the precision of the 16MAR30 results.

There are two limitations to the accuracy of the results reported here: no phase calibration was used at the KOKEE antenna, and no measurement of likely gravitational deformation of the antennas has been made. Analysis of the KOKEE phasecal delays from other sessions indicates that not using the phasecal may introduce a height error of as much as 2 mm. Gravitational effects on the KOKEE antenna may result in change in the measured height by as much as 10 mm, most likely in the opposite direction.

### 9.1 Improvements for future observing

The four KT sessions were scheduled before full operation was achieved for KOKEE12M. Observations in almost the entire northwest quadrant were excluded for this first set of measurements based on a preliminary visual estimate of the obscuration by KOKEE. As part of the experience subsequent to those observations, the effect of blockage by buildings, trees, towers, and antennas was assessed by measurement of the increase in system temperature as the antenna scanned the sky systematically, and a better horizon mask has been defined, thus enabling a better geometry for later observing sessions.

A primary conclusion of the VLBI2010 study (Petrachenko et al. 2009) was that station location precision increases as the number density (scans per hour) increases, with continued improvement down to at least four scans per minute. The low slew rates of KOKEE prohibit achieving that scan rate, but a higher rate than for the KT sessions could be achieved by reducing the minimum scan length, which was set at 30 seconds to be consistent with the on-going VGOS-only sessions, and by more careful consideration of the sources that are used.

Finally, a very important improvement can be implemented by turning on the phasecal at KOKEE and using the notch filter capability now available in *fourfit* to attenuate the cross-correlation of phasecal signals at the two antennas. This will enable correction of variations in delay in the signal chain from the receiver to the control room that may be introducing a bias in the height.

Thus, there are several changes that can be made for future observing sessions that will both improve the accuracy of the tie results and make it possible to achieve sub-millimeter precision in shorter sessions. This might enable more frequent measurement of the tie, thus allowing an evaluation of the stability of the two antennas.

## 9.2  Importance of structural deformation measurements and corrections

As the goal of 1 mm accuracy for the ITRF is approached, effects that have been ignored no longer can be. While gravitational deformation, which can cause an error of 10 mm or more, is likely to have a constant effect that can be applied retroactively once determined for an antenna, thermal deformation at a level of 1 mm or more varies through the year and even within a day. For an annual variation of 25°C, which is not uncommon among the instruments of the global networks that contribute to the ITRF, the thermal deformation for a 20 m antenna like KOKEE will change by about 2.5 mm. For a 12 m antenna like KOKEE12M the change is only 0.3 mm. However, for an optical survey to the intersection of axes, the changes in height for the extremes in temperature are correspondingly 4.1 mm and 1.9 mm, which if uncorrected will introduce errors of up to ±2 mm about the mean. The thermal deformation correction for a GNSS antenna mounted on a tower, if uncompensated, can be close to that of a VLBI antenna. To avoid systematic differences, it is especially important to use a common reference temperature for all techniques where they are co-located, as well as for any optical surveys at that location.

At the millimeter level of accuracy for position and 0.1 mm/yr for rate, long-term tilting of the antenna structure may be significant, as suggested by the pointing measurements of the KPGO antennas. Thus, monitoring the position of the antenna reference point relative to a set of local benchmarks is as important as other calibrations and corrections.

The largest uncertainty in the vector tie is the magnitude of the gravitational deformation of KOKEE. This could be reduced by almost an order of magnitude to the millimeter level if a careful survey were made.

## 9.3  Summary

We have measured the vector baseline between the legacy S/X antenna and the new VGOS antenna at Kokee Park Geophysical Observatory using VLBI phase delay observations. While the precision of these measurements can be sub-millimeter for a single 24-hr session, systematic errors due primarily to structural deformation increase the uncertainty to the level of a few millimeters.

It should be apparent from the results presented that, at the level of 1 mm, the accuracy of the ITRF is critically dependent on implementing the accounting for thermal deformation, obtaining the measurements of gravitational deformation for all antennas used in the construction of the ITRF, and monitoring local motions of the antennas.

The lessons learned from this study can be profitably applied to all co-located VGOS-legacy VLBI sites, thus providing a more global tie between the two systems. It would be beneficial to have these results prior to the analysis for ITRF2020 by which time there will be a several-year accumulation of VGOS results.

## Acknowledgements

We dedicate this work to Jim Long (deceased), NASA/Goddard Space Flight Center (GSFC), who helped with discussions regarding both the radio and optical ties at KPGO.

We greatly appreciate the efforts of Nancy Kotary, MIT Haystack Observatory, in developing the figure graphics.

We thank Ed Himwich, NVI, Inc., for obtaining the system temperature measurements and deriving the horizon mask for KOKEE12M and for providing the pointing data for the two antennas from which the tilt rates were derived.

Sergei Bolotin, NVI, Inc., has been very responsive in answering questions and implementing changes in *nuSolve*.

We appreciate discussions with Steven Breidenbach and Ben Erickson, NOAA/NGS, regarding the optical survey results at KPGO.

Axel Nothnagel, TUVienna, provided useful input on the application of the thermal deformation model.

Comments from two anonymous reviewers helped improve the manuscript.

**Declarations**
The corresponding author declares that he had full access to all the data in the study and takes responsibility for the integrity of the data and the accuracy of the data analysis.

Funding: The work by MIT Haystack Observatory was supported under NASA contracts NNG15HZ35C and 80GSFC20C0078.

Conflicts of interest/Competing interests: The authors state that there are no known conflicts of interest.

Data availability statement: The datasets generated and/or analyzed during the current study were obtained either as part of the commissioning phase of the KPGO antenna in 2016 or from a VLBI observing session in 2018, respectively. These datasets are publicly available through space geodesy data servers, such as the CDDIS, and from the corresponding author on reasonable request.

Code availability: The data were analyzed and figures were prepared using author-generated matlab™ scripts. The geodetic analysis used the publicly-available program *nuSolve* (Bolotin et al. 2019).

Authors' contribution statement:
AN, PE, GR, and CR designed, scheduled, and conducted the observations;
MT and RC correlated the observations;
MT, RC, and JB performed post-correlation processing;
AN, BC, PE, and DM performed the geodetic analysis;
AN, PE, and BC wrote the manuscript with input from all authors.

## Appendix: Thermal deformation information

The models for describing the various antenna types utilized in the IVS network are defined by Nothnagel (2009). For the KPGO antennas, which are both of the azimuth-elevation type, the model is:

$$\Delta\tau_{therm.i} = \frac{1}{c}\begin{pmatrix} \gamma_f\left(T(t-\Delta t_f)-T_0\right)\left(h_f \sin\varepsilon\right) \\ +\gamma_a\left(T(t-\Delta t_a)-T_0\right)\left(h_p \sin\varepsilon + AO\cos\varepsilon + h_v - F_a h_s\right) \end{pmatrix} \quad (A.1)$$

where

subscript *f* means *foundation* or, more appropriately for usage, *foundation material*
subscript *a* means *antenna* or, more appropriately for usage, *antenna material*
T = ambient temperature
$T_0$ = reference temperature
t = epoch of observation
$\gamma_f$ = coefficient of expansion of foundation material; default: $10 \times 10^{-6}$ per °C for concrete
$\gamma_a$ = coefficient of expansion of antenna material; default: $12 \times 10^{-6}$ per °C for steel
$\Delta t_f$ = temperature lag, foundation material
$\Delta t_a$ = temperature lag, antenna material
$h_f$ = dimension of foundation material plus pedestal, if concrete
$h_p$ = height of pedestal, if same material as antenna material
$h_v$ = dimension elevation axis to vertex of antenna
$h_s$ = distance from vertex of antenna to subreflector
$\varepsilon$ = elevation
AO = axis offset
$F_a$ = antenna focus factor: 1.8 for secondary focus; 0.9 for prime focus antenna

A diagram depicting the structural components of the KPGO antennas, based on Fig. 1 in Nothnagel (2009), is shown as Fig. 4.

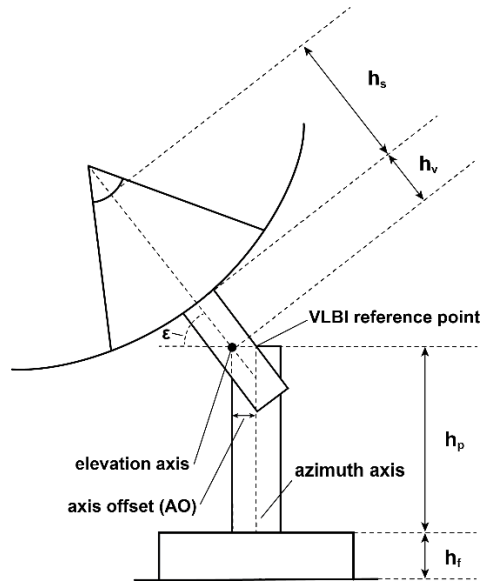

**Fig. 4** Alt-azimuth antenna mount with positive axis offset (adapted from Nothnagel (2009))

The reference temperature, $T_0$, is obtained from the Global Pressure and Temperature (GPT) model (Böhm et al. 2007) using an epoch of MJD 44357.3125. The delay (or height) correction should be subtracted from the measured value to adjust the position to that which would be observed at the reference temperature.

================================================================================

Table A1. Reference temperature and structural information from
https://raw.githubusercontent.com/anothnagel/antenna-info/master/antenna-info.txt.

|  | KOKEE | KOKEE12M |
|---|---|---|
| $T_0$ (°C): | 16.90 | 16.90 |
| $\gamma_f$ | $1.0*10^{-5}$ | $1.0*10^{-5}$ |
| $\gamma_a$ | $1.2*10^{-5}$ | $1.2*10^{-5}$ |
| $h_f$ (m) | 5.490 | 0.000 |
| $h_p$ (m) | 9.19 | 6.30 |
| $h_v$ (m) | 2.44 | 2.50 |
| $h_s$ (m) | 8.600 | 4.262 |
| AO (m) | 0.518 | 0.002 |
| $F_a$ (m) | 0.90 | 1.80 |

================================================================================

================================================================================

Table A2. Thermal deformation calculations for VLBI sessions (units are mm). Column headings are as given above for the antenna structure components. $h_{CT}$ is the total correction for each antenna. Total is the difference of the antenna totals in the sense KOKEE12M minus KOKEE.

| Session | T_mean (°C) | KOKEE | | | | | KOKEE12M | | | | | KOKEE12M-KOKEE |
|---|---|---|---|---|---|---|---|---|---|---|---|---|
|  |  | $h_F$ | $h_P$ | $h_V$ | $h_S$ | $h_{CT}$ | $h_F$ | $h_P$ | $h_V$ | $h_S$ | $h_{CT}$ | Total |
| 16MAR11 | 13.2 | -0.20 | -0.41 | -0.11 | 0.34 | -0.38 | 0.0 | -0.28 | -0.11 | 0.34 | -0.05 | 0.33 |
| 16MAR18 | 15.0 | -0.10 | -0.21 | -0.06 | 0.18 | -0.19 | 0.0 | -0.14 | -0.06 | 0.18 | -0.03 | 0.17 |
| 16MAR24 | 13.8 | -0.17 | -0.34 | -0.09 | 0.29 | -0.31 | 0.0 | -0.23 | -0.09 | 0.29 | -0.04 | 0.27 |
| 16MAR30 | 15.5 | -0.08 | -0.15 | -0.04 | 0.13 | -0.14 | 0.0 | -0.11 | -0.04 | 0.13 | -0.02 | 0.12 |

================================================================================

================================================================================

Table A3. Thermal deformation calculations for 2018 NGS survey (height units mm)

| Station | $T_{NGS}$ | $h_F$ | $h_P$ | $h_{CT}$ |
|---|---|---|---|---|
| KOKEE | 20.5 | 0.20 | 0.40 | 0.59 |
| KOKEE12M | 23.5 | 0.00 | 0.50 | 0.50 |

================================================================================

Thus, for the NGS 2018 survey the height correction to the reference temperature for KOKEE12M relative to KOKEE is -0.10 mm.